\documentstyle[prl,aps,epsf]{revtex}
\begin{document}

\title{\bf Influence of nonuniform critical current density profile
on magnetic field behavior of AC susceptibility in 2D Josephson
Junction Arrays}

\author{F.M. Araujo-Moreira$^{a}$, W. Maluf$^{a}$ and
S. Sergeenkov$^{a,b}$\footnote{Corresponding author (E-mail:
ssa@thsun1.jinr.ru; phone: 7-096-21-62123; fax: 7-096-21-65084)}}

\address {$^{a}$ Departamento de F\'{i}sica e Engenharia
F\'{i}sica, Grupo de Materiais e Dispositivos,\\ Centro
Multidisciplinar para o Desenvolvimento de Materiais
Cer\^amicos,\\
Universidade Federal de
S\~ao Carlos, S\~ao Carlos, SP, 13565-905 Brazil\\
$^{b}$ Bogoliubov Laboratory of Theoretical Physics, Joint
Institute for Nuclear Research,\\ Dubna 141980, Moscow Region,
Russia }
\address{\em (\today)}
\preprint \draft \maketitle
\begin{abstract}
Employing mutual-inductance measurements we study the magnetic
field dependence of complex AC susceptibility of artificially
prepared highly ordered (periodic) two-dimensional Josephson
junction arrays of unshunted $Nb-AlO_x-Nb$ junctions. The observed
behavior can be explained assuming single-plaquette approximation
of the overdamped model with an inhomogeneous critical current
distribution within a single junction.

\end{abstract}


\section{Introduction}

Despite the fact that Josephson Junction Arrays (JJA) have been
actively studied for decades, they continue to contribute to the
variety of intriguing and peculiar phenomena. To name just a few
recent examples, it suffice to mention the so-called paramagnetic
Meissner effect and related reentrant temperature behavior of AC
susceptibility, observed both in artificially prepared JJA and
granular superconductors (for recent reviews on the subject
matter, see, e.g.~\cite{1,2,3,4} and further references therein).
So far, most of the investigations have been done assuming an
ideal (uniform) type of array. However, it is quite clear that,
depending on the particular technology used for preparation of the
array,  any real array will inevitably possess some kind of
non-uniformity, either global (related to a random distribution of
junctions within array) or local (related to inhomogeneous
distribution of critical current densities within junctions). For
instance, recently a comparative study of the magnetic remanence
exhibited by disordered (globally nonuniform) 3D-JJA in response
to an excitation with an AC magnetic field was presented~\cite{5}.
The observed temperature behavior of the remanence curves for
arrays fabricated from three different materials ($Nb$,
$YBa_2Cu_3O_7$ and $La_{1.85}Sr_{0.15}CuO_4$) was found to follow
the same universal law regardless of the origin of the
superconducting electrodes of the junctions which form the array.

In the present paper, through an experimental study of complex AC
magnetic susceptibility $\chi (T,h_{AC})$ of the periodic
(globally uniform) 2D-JJA of unshunted $Nb-AlO_x-Nb$ junctions, we
present evidence for existence of the local type non-uniformity in
our arrays. Specifically, we found that in the mixed state region
$\chi (T,h_{AC})$ can be rather well fitted by a single-plaquette
approximation of the overdamped 2D-JJA model assuming a nonuniform
(Lorentz-like) distribution of the critical current density within
a single junction.

\section{Experiment}

Our samples consisted of $100\times 150$ unshunted tunnel
junctions. The unit cell had square geometry with lattice spacing
$a=46\mu m$ and a junction area of $5\times 5 \mu m^2$. The
critical current density for the junctions forming the arrays was
about $600 A/cm^2$ at $4.2K$, giving thus $I_C =150 \mu A$ for
each junction.

We used the screening method~\cite{6} in the reflection
configuration to measure the complex AC susceptibility $\chi =\chi
'+\chi ''$ of our 2D-JJA (for more details on the experimental
technique and set-ups see~\cite{7,8,9} ). Fig.1 shows the obtained
experimental data for the complex AC susceptibility $\chi
(T,h_{AC})$ as a function of the excitation field $h_{AC}$ for a
fixed temperature below $T_C$. As is seen, below $50mOe$ (which
corresponds to a Meissner-like regime with no regular flux present
in the array) the susceptibility, as expected, practically does
not depend on the applied magnetic field, while in the mixed state
(above $50mOe$) both $\chi '(T,h_{AC})$ and $\chi ''(T,h_{AC})$
follow a quasi-exponential field behavior of the single junction
Josephson supercurrent (see below).

\section{Discussion}

To understand the observed behavior of the AC susceptibility, in
principle one would need to analyze the flux dynamics in our
overdamped, unshunted JJA. However, given a well-defined (globally
uniform) periodic structure of the array, to achieve our goal it
is sufficient to study just a single unit cell (plaquette) of the
array. (It is worth noting that the single-plaquette approximation
proved successful in treating the temperature reentrance phenomena
of AC susceptibility in ordered 2D-JJA~\cite{4,7,8} as well as
magnetic remanence in disordered 3D-JJA~\cite{5}.) The unit cell
is a loop containing four identical Josephson junctions. Since the
inductance of each loop is ${\cal L}=\mu _0a\simeq 64 pH$ and the
critical current of each junction is $I_C =150 \mu A$, for the
mixed-state region (above $50mOe$) we can safely neglect the
self-field effects because in this region the inductance related
flux $\Phi _{L}(t)={\cal L}I(t)$ (here $I(t)$ is the total current
circulating in a single loop~\cite{10}) is always smaller than the
external field induced flux $\Phi _{ext}(t)=B_{ac}(t)S$ (here
$S\simeq a^2$ is the projected area of a single loop, and
$B_{ac}(t)=\mu _0h_{AC}\cos \omega t$ is an applied AC magnetic
field). Besides, since the length $L$ and the width $w$ of each
junction in our array is smaller than the Josephson penetration
depth $\lambda _J=\sqrt{\Phi _0/2\pi \mu _0dj_{c0}}$ (where
$j_{c0}$ is the critical current density of the junction, $\Phi
_0$ is the magnetic flux quantum, and $d=2\lambda _{L}+\xi$ is the
size of the contact area with $\lambda _{L}(T)$ being the London
penetration depth of the junction and $\xi $ an insulator
thickness), namely $L\simeq w \simeq 5\mu m$ and $\lambda _J\simeq
20\mu m$ (using $j_{c0}=600A/cm^2$ and $\lambda _{L}=39nm$ for
$Nb$ at $T=4.2K$), we can adopt the small-junction
approximation~\cite{10} for the gauge-invariant superconducting
phase difference across the $i$th junction (by symmetry we assume
that~\cite{7,8} $\phi _1=\phi _2=\phi _3=\phi _4\equiv \phi _i$)
\begin{equation}
\phi _i(x,t)=\phi _0+\frac{2\pi B_{ac}(t)d}{\Phi _0}x
\end{equation}
where $\phi _0$ is the initial phase difference.

The net magnetization of the plaquette is $M(t)=SI_s(t)$ where the maximum supercurrent
(corresponding to $\phi _0=\pi /2$) through an inhomogeneous Josephson contact reads
\begin{equation}
I_s(t)=\int\limits_0^Ldx\int\limits_0^wdyj_c(x,y)\cos \phi _i(x,t)
\end{equation}
For the explicit temperature dependence of the Josephson critical current density
\begin{equation}
j_{c0}(T)=j_{c0}(0)\left [\frac{\Delta (T)}{\Delta (0)}\right ]\tanh \left [\frac{\Delta
(T)}{2k_BT}\right ]
\end{equation}
we used the well-known~\cite{11} analytical approximation for the
BCS gap parameter (valid for all temperatures), $\Delta (T)=\Delta
(0)\tanh \left (2.2 \sqrt{\frac{T_C-T}{T}}\right )$ with $\Delta
(0)=1.76k_BT_C$.

In general, the values of $\chi '(T,h_{AC})$ and $\chi
''(T,h_{AC})$ of the complex harmonic susceptibility are defined
via the time-dependent magnetization of the plaquette as follows:
\begin{equation}
\chi '(T,h_{AC})=\frac{1}{\pi h_{AC}}\int_0^{2\pi}d(\omega t)\cos
(\omega t) M(t)
\end{equation}
and
\begin{equation}
\chi ''(T,h_{AC})=\frac{1}{\pi h_{AC}}\int_0^{2\pi}d(\omega t)\sin
(\omega t) M(t)
\end{equation}
Using Eqs. (1)-(5) to simulate the magnetic field behavior of the
observed AC susceptibility of the array, we found that the best
fit through all the data points and for all temperatures is
produced assuming the following nonuniform distribution of the
critical current density within a single junction~\cite{10}
\begin{equation}
j_c(x,y)=j_{c0}(T)\left (\frac{L^2}{x^2+L^2}\right )\left (\frac{w^2}{y^2+w^2}\right )
\end{equation}
It is worthwhile to mention that in view of Eq.(2), in the
mixed-state region the above distribution leads to approximately
exponential field dependence of the maximum supercurrent
$I_s(T,h_{AC})\simeq I_s(T,0)e^{-h_{AC}/h_0 }$ which is often used
to describe critical-state behavior in type-II
superconductors~\cite{12}. Given the temperature dependencies of
the London penetration depth $\lambda _L(T)$ and the Josephson
critical current density $j_{c0}(T)$, we find $h_0(T)=\Phi _0/2\pi
\mu _0\lambda _J(T)L\simeq h_0(0)(1-T/T_C)^{1/4}$ for the
temperature dependence of the characteristic field near $T_C$.
This explains the improvement of our fits (shown by solid lines in
Fig.1) for high temperatures because with increasing the
temperature the total flux distribution within a single junction
becomes more regular which in turn validates the use of the
small-junction approximation.

In summary, we found clear experimental evidence for the influence
of the junction nonuniformity on magnetic field penetration into
the periodic 2D array of unshunted Josephson junctions. By using
the well-known AC magnetic susceptibility technique, we have shown
that in the mixed-state regime the AC field behavior of the
artificially prepared array is reasonably well fitted by the
single-plaquette approximation of the overdamped model of 2D-JJA
assuming inhomogeneous (Lorentz-like) critical current
distribution within a single junction.

\acknowledgements

We thank  P. Barbara, C.J. Lobb, and R.S. Newrock for useful
discussions.  S.S. and F.M.A.M. gratefully acknowledge financial
support from Brazilian Agency FAPESP under grant 2003/00296-5.

\newpage
\begin{figure}[!h]
\begin{center}
\vspace{2cm} \epsfysize=6.5cm \epsffile{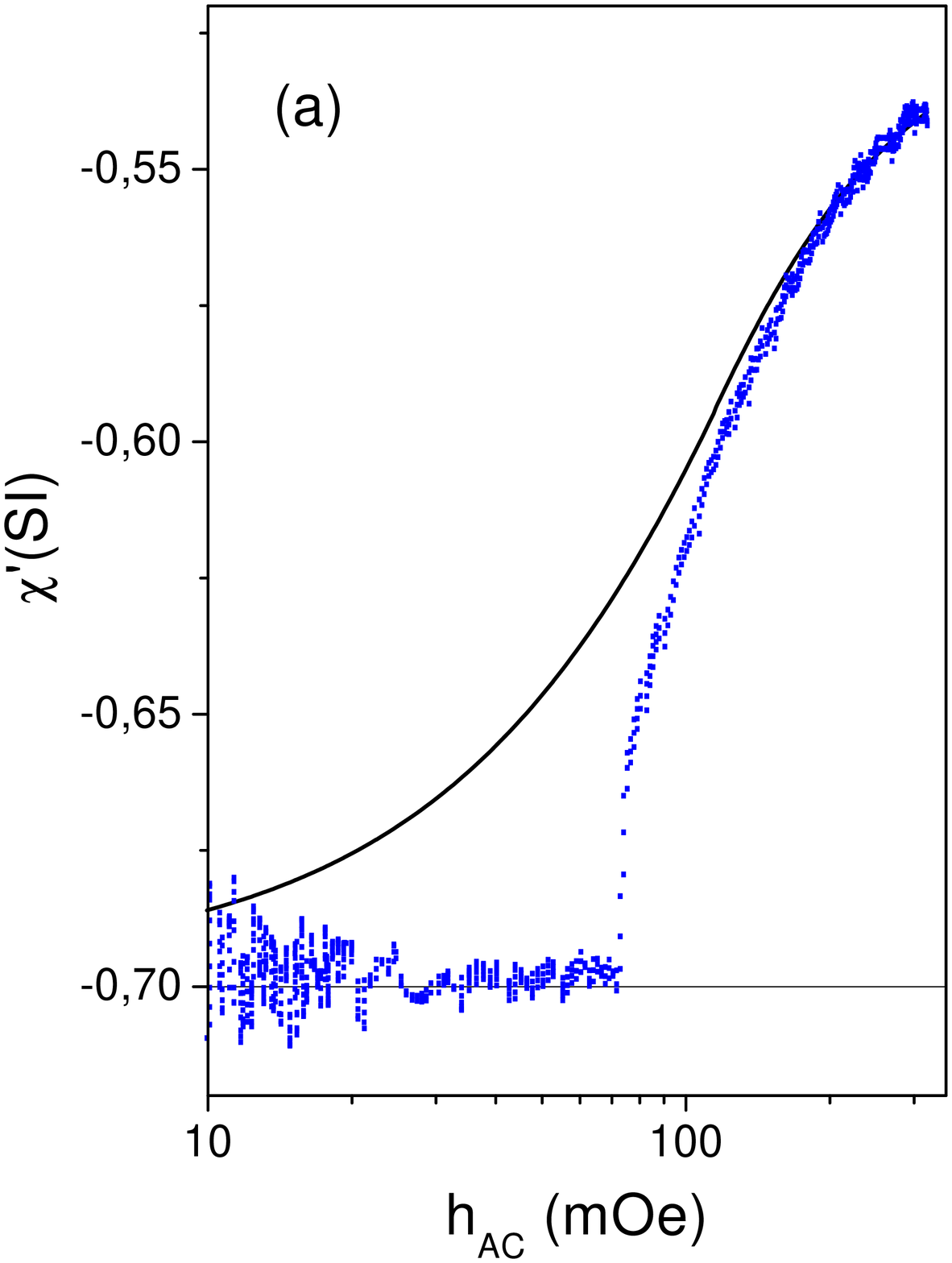} \hspace{0.5cm}
\epsfysize=6.5cm \epsffile{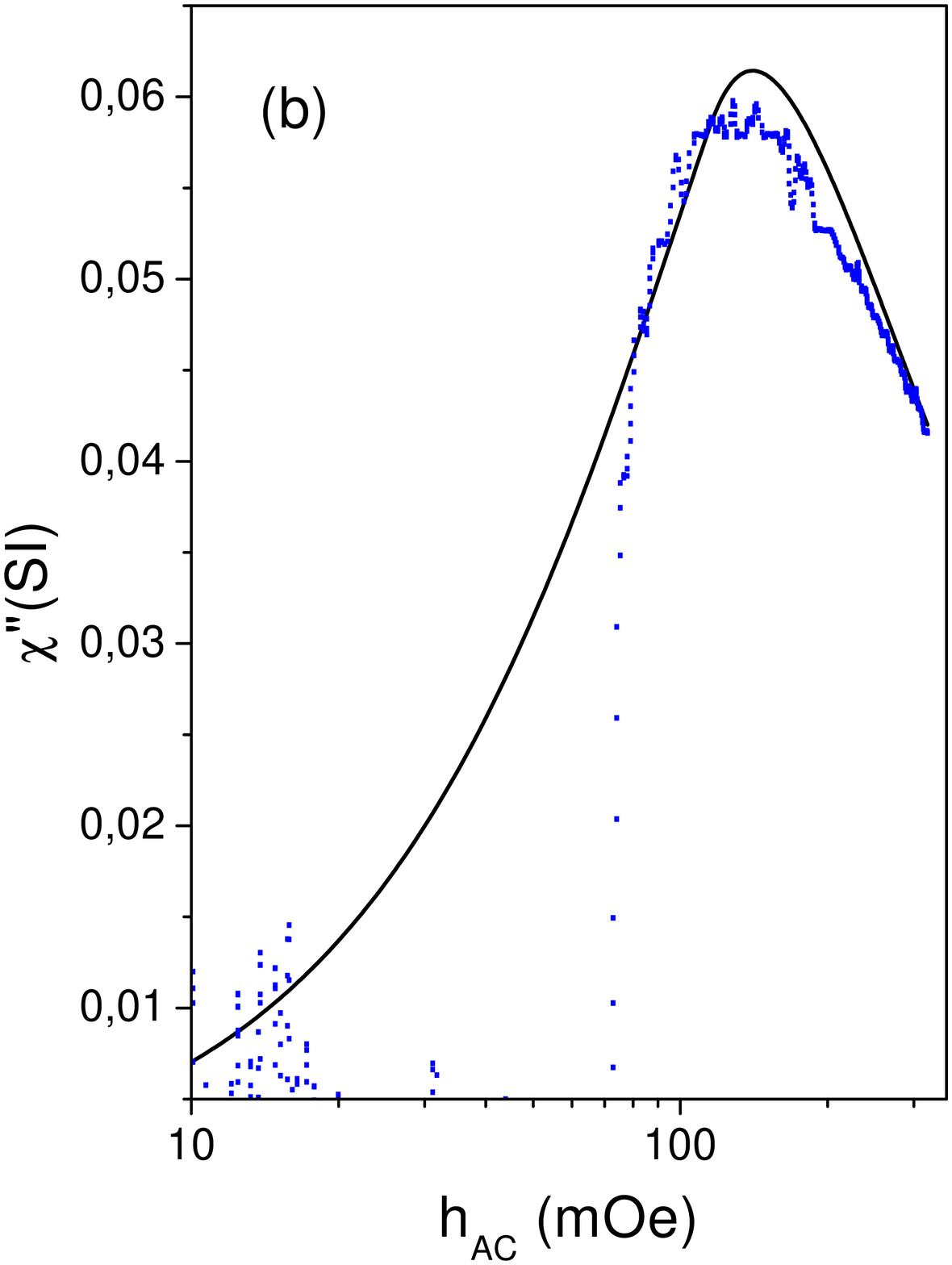} \hspace{0.5cm}
\epsfysize=6.5cm \epsffile{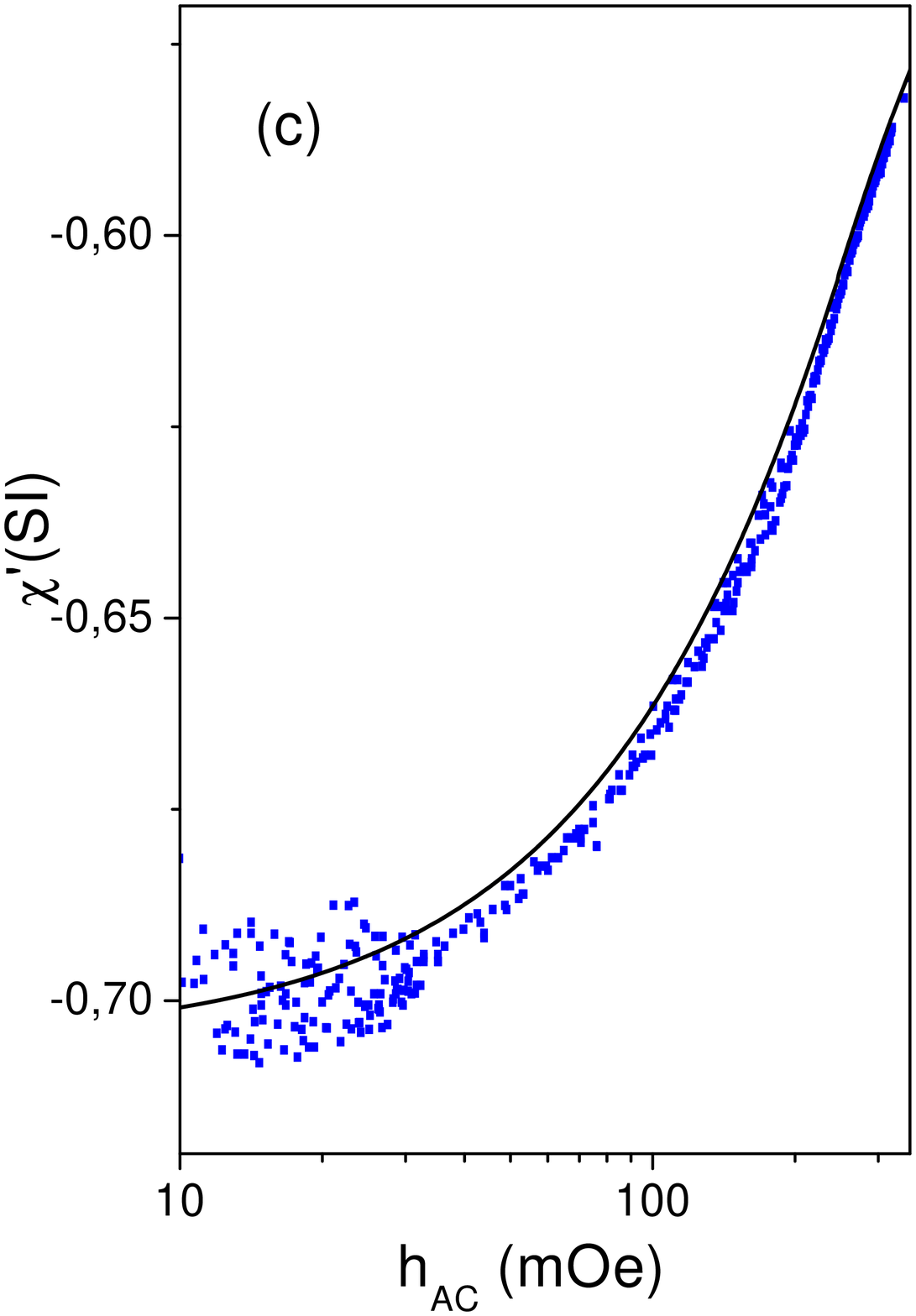} \\ \vspace{1cm}
\epsfysize=6.5cm \epsffile{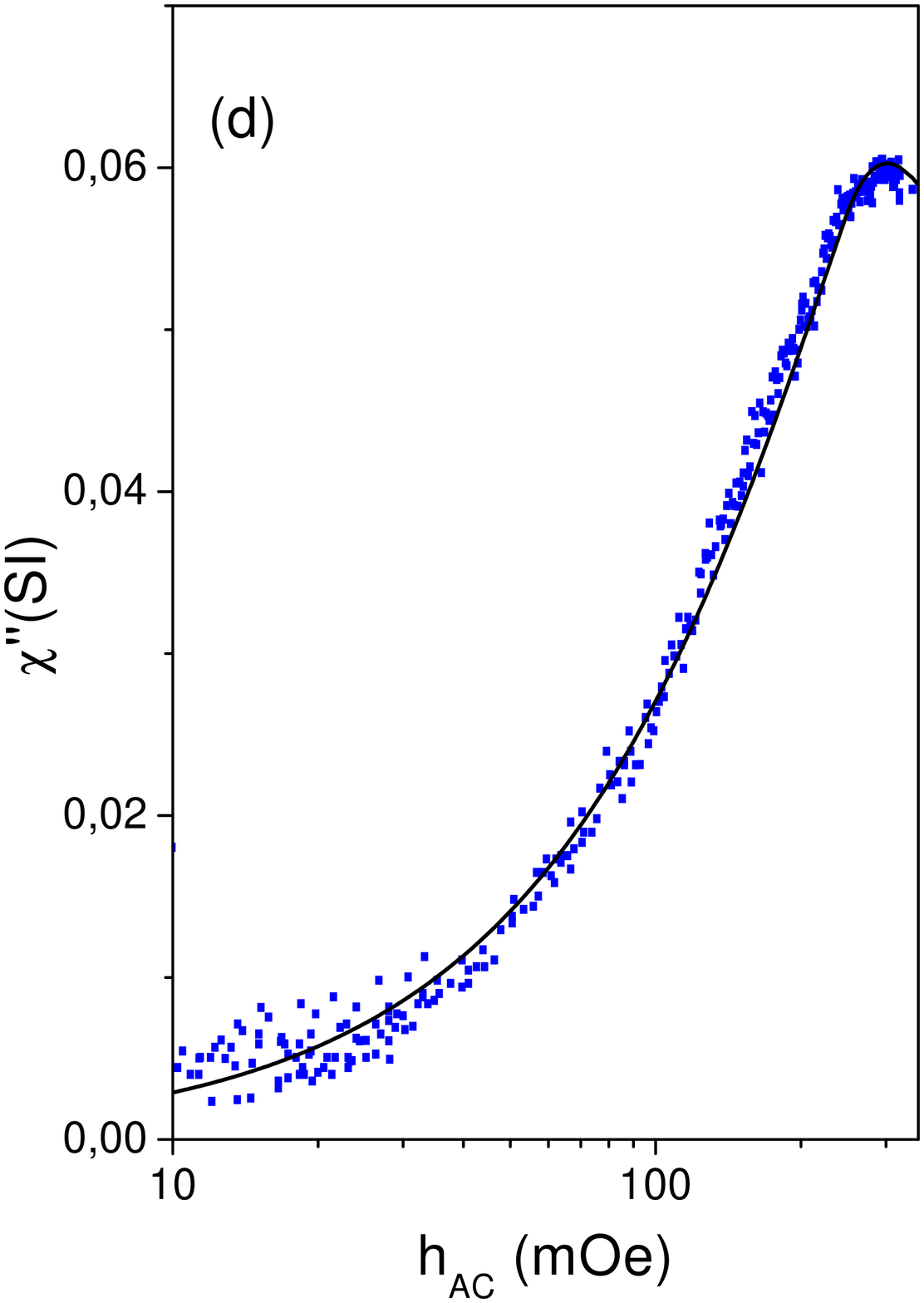} \hspace{0.5cm}
\epsfysize=6.5cm \epsffile{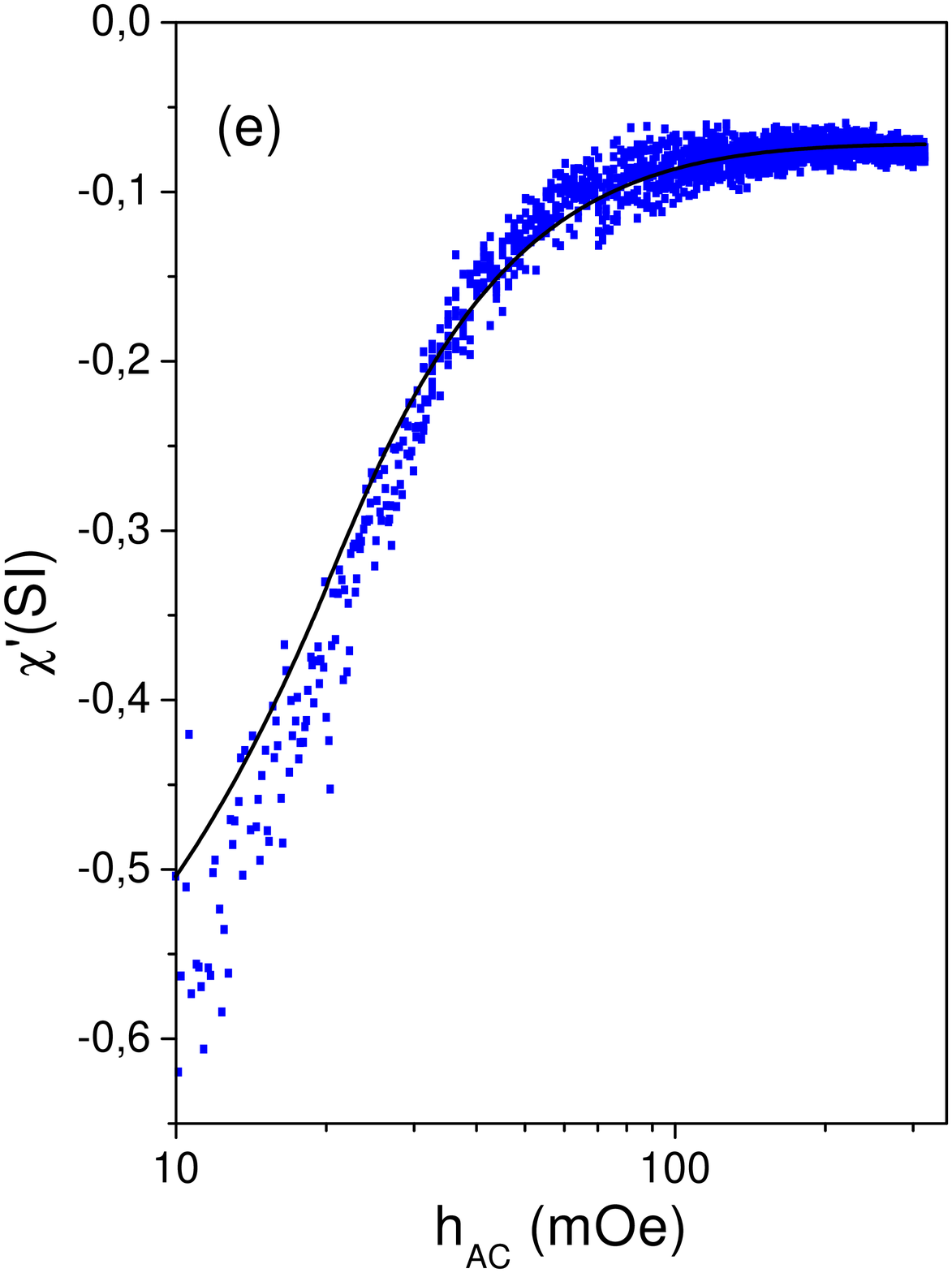} \hspace{0.5cm}
\epsfysize=6.5cm \epsffile{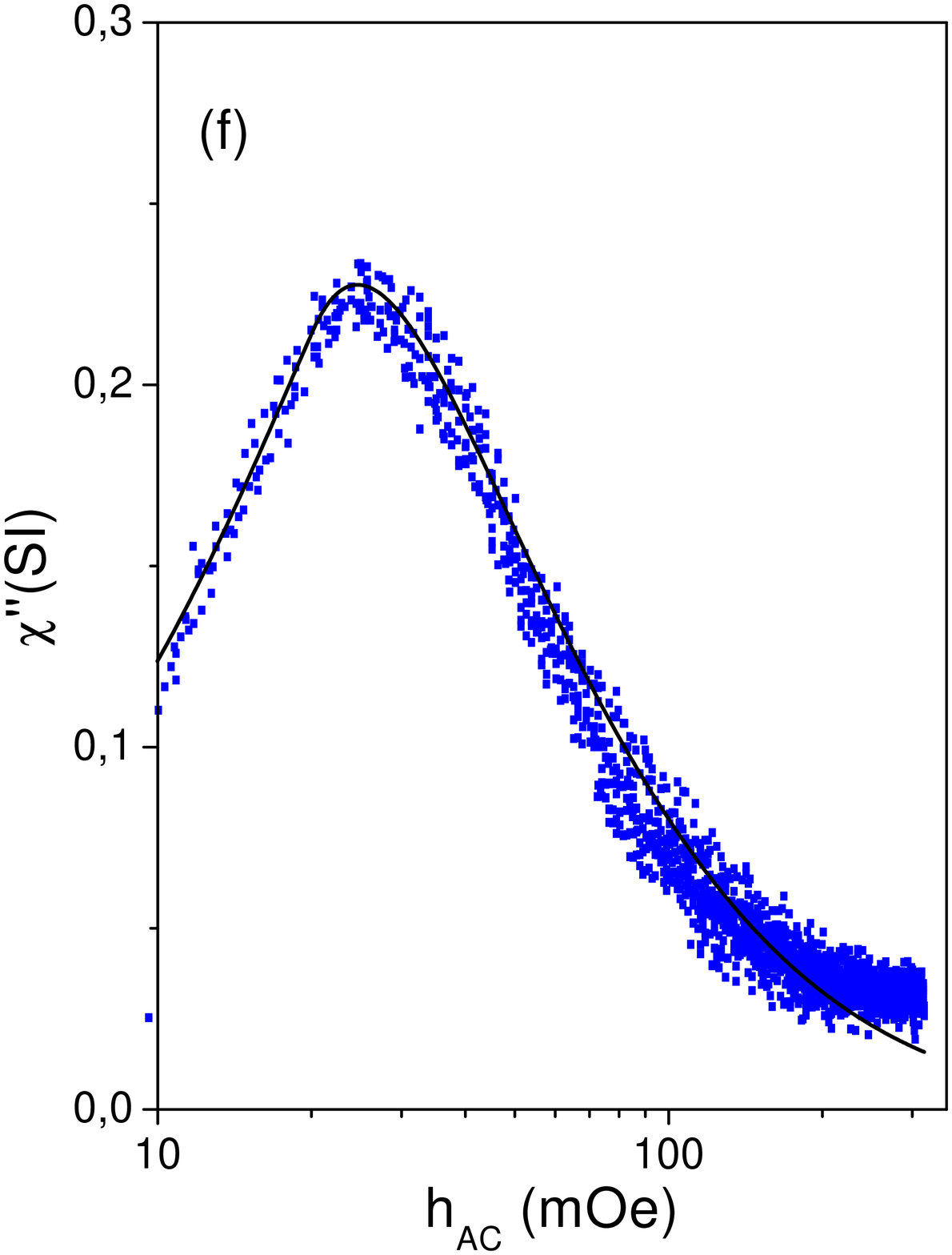} \vspace{2cm} \caption[]{{The
dependence of the magnetic susceptibilities, $\chi '(T,h_{AC})$
and $\chi ''(T,h_{AC})$, on AC magnetic field amplitude $h_{AC}$
for different temperatures:  $T = 4.2 K$ (a,b), $T = 6 K$ (c,d),
and $T = 8 K$ (e,f). Solid lines correspond to the fitting of the
2D-JJA model with nonuniform critical current profile for a single
junction (see the text).}}
\end{center}
\end{figure}

\end{document}